\documentclass[multphys,vecphys]{svmult}

\usepackage{makeidx}     
\usepackage{graphicx}    
\usepackage{multicol}    

\makeindex             

\begin{document}

\hyphenation{ele-ments}

\title*{ESC Observations of SN~2002er around Maximum Light}
\author{G. Pignata\inst{1}, F. Patat\inst{1},
R. Kotak\inst{2}, P. Meikle\inst{2}, M. Stritzinger\inst{3}, W. Hillebrandt\inst{3} }
\institute{European Southern Observatory, Karl-Schwarzschild str. 2, D-85748 Garching bei M\"unchen, Germany.
\texttt{gpignata@eso.org}
\and Imperial College London, Prince Consort Road, London SW7 2BZ, United Kingdom.
\and Max-Planck-Institute f\"ur Astrophysik, Karl-Schwarzschild str. 1, D-85741 Garching bei M\"unchen, Germany.}
%
%
\maketitle

\abstract

We present preliminary results of the observational campaign on SN~2002er carried out by the European Supernova Collaboration (ESC). 
For this supernova the optical coverage from epochs $-$11 days to +50 days  is excellent, both in imaging and spectroscopy. The extinction deduced from the photometry is $E(B-V)=0.35$, while $\Delta m_{15}=1.31$.
In general SN~2002er behaves like a ``normal'' type Ia supernova, very similar to  SN~1996X.

\section{Introduction }
\label{sec:1}
In order to carry out a comprehensive study of the physics of Type Ia
Supernovae (SNeIa), all the major European institutes working in this
field have formed the European Supernova Collaboration (ESC). This is
partially supported as an EU European Research Training Network,
commencing operation in 2002.  The ESC has already carried out
detailed monitoring of a number of nearby SNe Ia using a large number
of telescopes and instruments.  In this poster we present preliminary
results for SN 2002er in UGC 10743, one of the first targets followed
by the ESC.\\
SN~2002er ($\alpha = 17^h 11^m 29^s.88$, 
$\delta = + 7^\circ 59' 44''.8$, J2000)  was discovered on August 23.2, IAU 7959\cite{Wood}.  On the basis of a low resolution spectrum taken at La Palma with the INT on August 26.9 UT, the SN candidate was classified by some of the ESC members as a type Ia, approximately 10 days before maximum brightness \cite{Smartt}.

\section{Interstellar Extinction}
\label{sec:2}
In order to estimate the intrinsic luminosity of the SN, it is crucial to estimate an accurate  value for the reddening and to correct for this effect. To compute $E(B-V)$, we have used different methods. Using the  Lira locus \cite{Phillps} in the tail phase, we get 0.44 $\pm$ 0.08. If we compare our observed $B-V$ colour around $B$ maximum with that given  by Phillips \cite{Phillps}, we obtain 0.31 $\pm$ 0.05. Finally considering the results of CMAGIC method \cite{Wang} we get  0.28 $\pm$ 0.04.  Independent estimates
of $E(B-V)$ will be provided by spectral  modelling of the spectroscopic data. For our purposes, here we adopt the weighted mean of the above mentioned values of reddening, i.e. $E(B-V)$=0.35.

\section{$UBVRI$ Light Curves}
\label{sec:3}

The $UBVRI$ light curves of SN~2002er are shown in Fig.~1. The $\Delta m_{15}$ of SN~2002er is 1.31 (Table 1).
For comparison, the light curves of other Type Ia SNe with similar values of $\Delta m_{15}$ are  also sketched.
As can be clearly seen from the figure, the best match is with SN~1996X.
This object resembles SN~2002er even in the I band, where the differences between SNe are usually more 
pronounced.

\begin{figure}
\centering
\includegraphics[height=8.5cm]{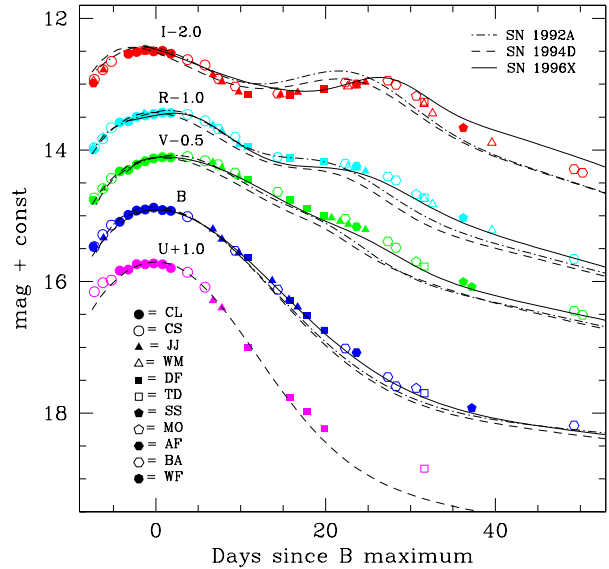}
%
%
\caption{$UBVRI$ light curves of SN~2002er, The ordinate scale refers to the $B$-band, The other bands are shifted by the amount  shown in the plot. Different symbols refer to different instruments with which the SN was observed. The dashed lines and dashed-dotted lines represent the $UBVRI$ light curve of 1994D ($\Delta m_{15}=1.32$), \cite{Patat}  and SN~1992A ($\Delta m_{15}=1.47$) \cite{Suntzeff}. The solid lines refer to the $BVRI$ light curve of  SN~1996X ($\Delta m_{15}=1.31$) \cite{Salvo}.}
\label{fig:1}       
\end{figure}

\noindent The main photometric parameters of SN~2002er, are summarized in the Table 1. The extinction corrected colors, decline rates, light curve shape and time offsets between maxima in 
different passbands  are typical of ``normal'' SNe.

\begin{table}
\centering
\caption{Main photometric parameters of SN~2002er.}
\label{tab:1}       
%
%
\begin{tabular}{cccccc}
\hline
\hline
Filter~ & ~mag. first max~ & ~phase~ &  ~$\Delta m_{15}$~ & ~mag.~ second max ~& ~phase\\
& & (days) & & & (days)\\
\hline
U  & 14.69 & -0.4 & - & - & -\\
B &  14.88  & 0.0 & 1.31 &- & -\\
V & 14.62 & 2.1 & -& - & -\\
R & 14.42  & 1.7 & - & - & -\\
I &14.45   & -0.2 & - & 14.93  & 25.5\\
\hline
\multicolumn{6}{l}{~~~~~~$(U-B)_0=$$-$0.63 ,~ $(B-V)_0=$$-$0.10, ~ $(V-I)_0=$$-$0.27}\\
\hline
\hline
\end{tabular}
\end{table}

\section{Absolute magnitude and Bolometric Light Curves}
\label{sec:4}

We can derive  the absolute magnitude of 
SN~2002er in two ways.\\
In the first case we assume $H_0$=71  km s$^{-1}$ Mpc$^{-1}$ \cite{Spergel}  and, from the radial velocity  corrected for LG 
infall onto Virgo ($v_r$=2694 km s$^{-1}$), we derive a distance modulus $\mu$=32.9$\pm$0.3 for host galaxy. 
Then, taking into account the estimated $A_B=4.315 \times E(B-V)= 1.47$ \cite{Schlegel} we obtain $M^{max}_{B}=-$19.5$\pm$0.3.\\
Alternatively, we can use the relation between $M^{max}_{B}$ and $\Delta m_{15}$ found by Hamuy \cite{Hamuy} 
This empirical law, applied to the case of SN~2002er, gives $M^{max}_{B}=-$19.1$\pm$ 0.3. 
We note that the two values agree to within the errors, although the associated errors are
rather large.

%
\begin{figure}
\centering
\includegraphics[height=8.5cm]{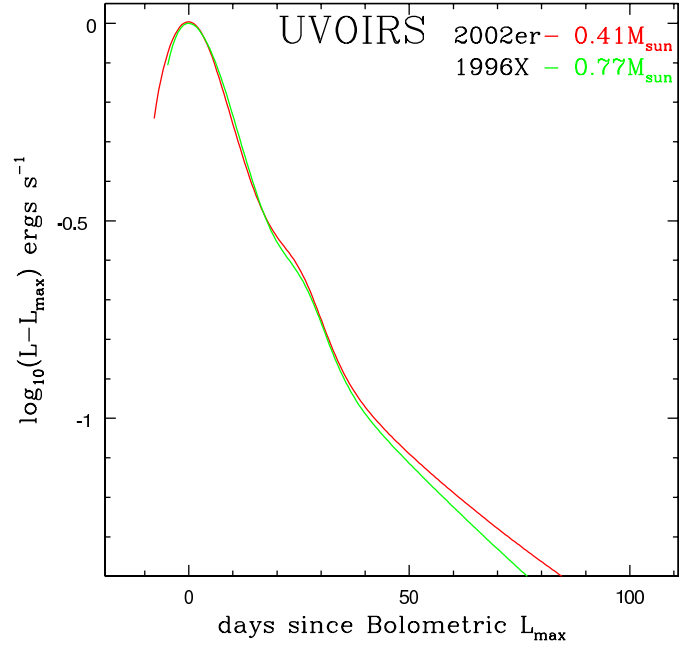}
%
%
\caption{Bolometric light curves of SN~2002er (red line) and SN~1996X (green line).}
\label{fig:1}       
\end{figure}
\noindent With our well sampled $UBVRI$ data of SN~2002er we are able to construct a $UVOIR$ light curve 
and obtain a reliable estimate of the total luminosity at maximum light.
The $UVOIR$ light curve, computed using the method described in Vacca \& Leibundgut \cite{Vacca}, is presented in Fig.~2.
For comparison we have overplotted the corresponding curve for SN~1996X \cite{Salvo, Riess}, which has a similar $\Delta m_{15}$ (1.31). Distance moduli were computed assuming $H_0$ as in the previous section. For extinction to SN~2002er we have used the  value discussed above, while for SN~1996X we have adopted the estimates available in the literature  $\mu$=32.7, $E(B-V)$=0.08. Using  Arnett's rule \cite{Arnett}, we obtain Nickel masses of 0.41 $M_{\odot}$ and 0.77 $M_{\odot}$  for SN~2002er and SN~1996X respectively. The similarity in the shape of the $UVOIR$ light curves during early epochs to beyond the $\sim$28-day feature is remarkable considering that the difference in maximum luminosity among normal SNe Ia is of the order a factor of 2.

\setcounter{table}{1}

\section{Spectral analysis}
\label{sec:5}

The spectral evolution of SN 2002er from $-$10$^d$ to +35$^d$ is shown in  Fig.~3
which presents just a subset of the entire dataset. The evolution is very similar to 
that of other normal type Ia supernovae like SN~1994D.

\begin{figure}
\centering
\includegraphics[height=8.5cm]{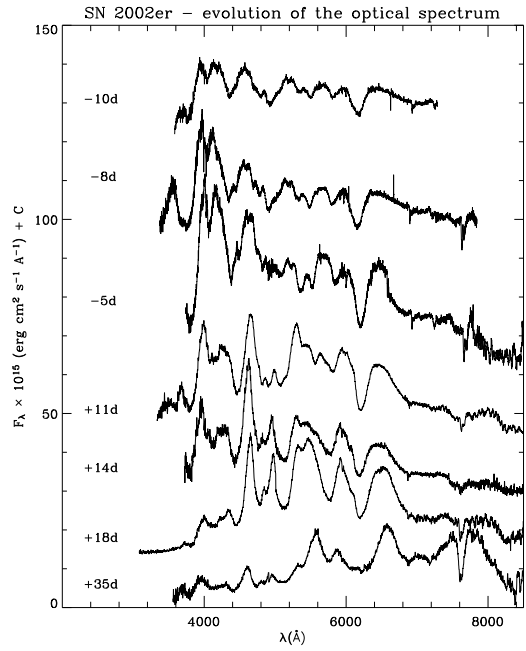}
%
%
\caption{Spectral evolution of SN~2002er.}
\label{fig:1}       
\end{figure}

\noindent The dominant features 
are due to the Fe-group and intermediate mass elements (notably Si, Ca, Mg). 
The lines exhibit the characteristic P-Cygni profiles, and the minima of the
absorption components shift to redder wavelengths (i.e. lower expansion
velocities) with time. This is most apparent in the SiII 6150 \AA\/ feature.
The blue edge of the SiII feature at $-$10$^d$ corresponds to a velocity of
$\sim$20500 km s$^{-1}$ which is comparable to that found for SN~1994D at a similar epoch ($\sim$21600 km s$^{-1}$, \cite{Patat} see Fig.~4). Note that the spectra in Fig ~3. have not been de-reddened.\\

\begin{figure}
\centering
\includegraphics[height=7cm]{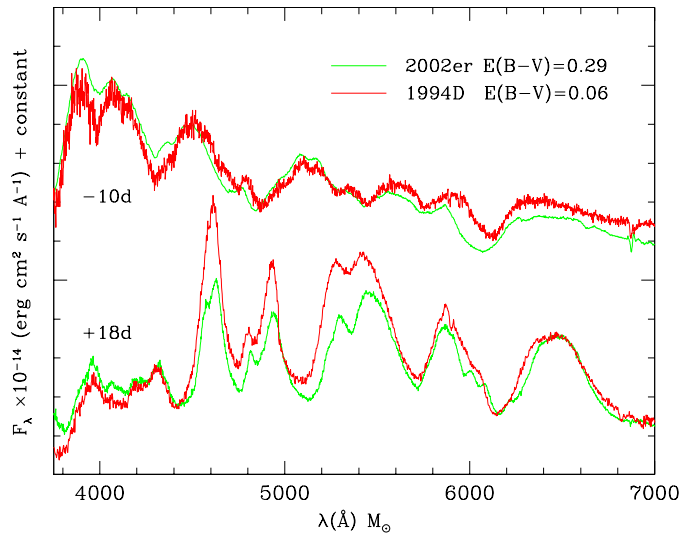}
%
%
\caption{Comparison of the spectra of SN~2002er and SN~1994D at two different epochs.}
\label{fig:1}       
\end{figure}

\noindent This work was supported by the EC
through contract HPRN-CT-2002-00303.

%
%
%
%
%

%
%



\printindex
\end{document}